\documentclass[copyright,creativecommons]{eptcs}
\providecommand{\event}{CompMod 2009} 
\providecommand{\volume}{x}
\providecommand{\anno}{2009}
\providecommand{\firstpage}{1}
\providecommand{\eid}{x}

\usepackage{url}
\usepackage{amssymb}
\usepackage{amsthm}
\usepackage{amsmath,amsfonts,amssymb}
\usepackage{mathrsfs}
\usepackage{pgf,pgfarrows,pgfnodes,pgfautomata,pgfheaps,pgfshade}
\usepackage{multicol}
\usepackage{color}

\theoremstyle{plain}

\newtheorem{proposition}{Proposition}[section]

\theoremstyle{definition}
\newtheorem{definition}{Definition}[section]
\newtheorem{example}{Example}[section]
\theoremstyle{remark}

\newtheorem{remark}{Remark}[section]

\newcommand{\Q}{\mathbb{Q}}
\newcommand{\bbR}{\mathbb{R}}

\newcommand{\eps}{\varepsilon}

\newcommand{\vr}[1]{\mathbf{#1}}

\newcommand{\calT}{\mathcal{T}}

\newcommand{\calA}{\mathcal{A}}

\newcommand{\calD}{\mathcal{D}}

\newcommand{\defeq}{\stackrel{\mbox{\tiny def}}{=}}

\newcommand{\sccptuple}[3]{ (A_{#2}^{#3},\calD_{#2}^{#3},\vr{#1_{#2}}, init_{#2}^{#3}(\vr{\vr{#1_{#2}}})) }

\newcommand{\TC}{\mathfrak{TC}}
\newcommand{\TD}{\mathfrak{TD}}
\newcommand{\TS}{\mathfrak{TS}}
\newcommand{\init}[1]{init_{#1}}

\newcommand{\TDHA}[1]{\mathcal{T}_{#1}}

\newcommand{\exit}[1]{\mathbf{e_1}[#1]}
\newcommand{\enter}[1]{\mathbf{e_2}[#1]}
\newcommand{\stoich}[1]{\mathbf{stoich}[#1]}
\newcommand{\rate}[1]{\mathbf{rate}[#1]}
\newcommand{\cmode}[1]{\mathbf{cmode}[#1]}
\newcommand{\priority}[1]{\mathbf{priority}[#1]}
\newcommand{\guard}[1]{\mathbf{guard}[#1]}
\newcommand{\reset}[1]{\mathbf{reset}[#1]}

\newcommand{\sCCP}{\textbf{sCCP}}

\newcommand{\xs}[3]{#1_#2,\ldots,#1_#3}

\newcommand{\mode}[2]{S_{#1}(#2)}

\begin{document}

\title{Hybrid Semantics of Stochastic Programs with Dynamic Reconfiguration}
\author{Luca Bortolussi
\institute{Dept. of Mathematics and Informatics,\\
University of Trieste, Italy.} \email{luca@dmi.units.it}
 \and Alberto Policriti
\institute{Dept. of Mathematics and Informatics,\\
University of Udine, Italy.} \institute{Istituto di Genomica
Applicata, Udine, Italy.} \email{policriti@dimi.uniud.it} }

\def\titlerunning{Hybrid semantics of Stochastic Programs with Dynamic Reconfiguration}
\def\authorrunning{L. Bortolussi and A. Policriti}

\maketitle

\begin{abstract}

We begin by reviewing a technique to approximate the dynamics of
stochastic programs---written in a stochastic process algebra---by a
hybrid system, suitable to capture a mixed discrete/continuous
evolution. In a nutshell, the discrete dynamics is kept stochastic
while the continuous evolution is given in terms of ODEs, and the
overall technique, therefore,  naturally associates  a Piecewise
Deterministic Markov Process with a stochastic program.

The specific contribution in this work consists in an increase of
the flexibility of the translation scheme, obtained by allowing  a
\emph{dynamic} reconfiguration of the degree of
discreteness/continuity of the semantics.

We also discuss the relationships of this approach with other
hybrid simulation strategies for biochemical systems.

\end{abstract}

\section{Introduction}\label{sec:introduction}

Models in systems biology tend to  cluster around two families of
mathematical tools: differential equations and stochastic processes.
Even though, physically speaking, stochastic models have firmer
grounds
\cite{SB:Gillespie:1976:gillespieAlgorithm,SB:Gillespie:1977:gillespieAlgorithm},
their computational analysis is much more costly than that of their
differential counterpart. In any case, ODE-based descriptions of
biological systems are often valuable and provide deep insights.
Indeed, it is known that, limiting to mass action models, ODE's are
an approximation of the average of stochastic models, and the
differences between the two vanish in the thermodynamic
limit~\cite{SB:Gillespie:2000:ChemicalLangevinEquation} (i.e. when
populations and system's size go to infinity). Recently, there have
been many attempts to mix these two techniques, at least as far as
simulation of biological systems is concerned, resulting in several
hybrid simulation
algorithms~\cite{SB:Wilkinson:2006:StochasticModellingSB,SB:Pahle:2009:HybridSimSurvey}.
Hybrid dynamical systems have also been a hot topic in the last two
decades, with much research work spanning across the boundary
between computer science and  engineering control. The best known
model among hybrid dynamical systems are \emph{hybrid
automata}~\cite{HA:Henziger:1996:SurveyHybridAutomata}. Stochastic
extensions of such concept are also receiving recently much
attention~\cite{HA:BujorianuLygeros:2004:GSHA}, although stochastic
hybrid systems have a somewhat longer
tradition~\cite{STOC:Davis:1993:PDMP}. In both cases, most of the
interest is in the development of automated reasoning tools rather
than in simulation.

It is widely recognized that Computational Systems Biology  can
highly benefit from modeling approaches embodying \emph{some}
stochastic ingredient. A very popular line along which such
incorporation is realized, is based on the use of stochastic process
algebras~\cite{SB:ShapiroRegev:2002:CellsAsComputation,SB:HillstonGilmoreCalder:2006:ERKPathwayPEPA},
which are proposed as front-end languages to (automatically)
generate mathematical models, usually Continuous Time Markov
Chains(CTMC), see~\cite{SB:Wilkinson:2006:StochasticModellingSB}.
Recently, such process algebra based languages have also been
endowed with semantics based on
ODE~\cite{PA:Hillston:2005:ODEandPEPA}, which increase the
flexibility of such tools.

Many proposals of hybrid simulation algorithms for systems of
biochemical reactions have been put
forward~\cite{SB:Rawlings:2002:HybridSimulation,
SB:Neogi:2004:HybridSimulationDynamicPartitioning,
SB:HaseltineRawlings:2005:OriginsApproximationChemKin,
SB:Kaznessis:2005:HybridSimulation,SB:Kiehl:2004:hybridSimulation,
SB:Griffith:2006:HybridSimulation,SB:Alfonsi:2005:HybridSimulation}.
Their salient feature is a description of one part of the system as
continuous, while keeping the other discrete and stochastic. The
basic idea is to find the best trade off between accuracy and
computational efficiency (stochastic simulations are much more
expensive than ODE simulation).

In this paper we continue a programme which aims to increase even
more the flexibility of stochastic process algebras by providing
them with a very general semantics based on (stochastic) hybrid
systems, encompassing CTMC and ODE as special cases. Such an
approach is motivated not only by the gain in flexibility, but also
by the goal of exploiting, in a systematic manner, automated
reasoning tools to provide as much information as possible from a
given model. Our stochastic process algebra of choice is stochastic
Concurrent Constraint Programming
(\sCCP)~\cite{SB:Bortolussi:2008:BiomodelingSCCP:Journal}, an
extension of CCP~\cite{PA:Saraswat:1990:CCP} in the stochastic
setting. In addition to the standard CTMC-based semantics, we have
also provided sCCP with an ODE-based
semantics~\cite{PA:Bortolussi:2007:SCCPandODEjournal} and with an
hybrid automata based semantics. Moreover, hybrid semantics has been
proposed both with a fixed or user-defined amount of continuously
approximated components
(see~\cite{SB:Bortolussi:2009:HybridSCCPstaticJournal,SB:Bortolussi:2009:HybridsCCPLattice}).

In this paper we  extend our work by introducing a semantics based
on \emph{Stochastic} Hybrid Automata, thereby guaranteeing the
possibility of parameterizing  the degree of continuity introduced
in the model. The approach allows also a \emph{dynamic}
reconfiguration of such degree, in accordance to properties of the
current state of the system. This allows the description in a formal
setting of different hybrid simulation strategies, opening the way
for their use in the context of process algebra modelling.

We will start our presentation by introducing, in
Section~\ref{sec:TDHA}, a high level description of the target
stochastic hybrid systems,  suitable to be easily mapped to the
well-established formalism of Piecewise Deterministic Markov
Processes (see supplementary material~\cite{supp}). The formalism
introduced in Section~\ref{sec:TDHA}, called Transition-Driven
Stochastic Hybrid Automata (TDSHA), will act as the intermediate
layer in the definition of the stochastic hybrid semantics of \sCCP.
Section~\ref{sec:sCCP} briefly introduces the \sCCP\ language, while
Section~\ref{sec:sCCPtoTDHA} presents the mapping from \sCCP\ to
TDSHA. Collections of TDSHAs can be organized in a lattice, whose
definition and basic properties are presented in
Section~\ref{sec:latticeTDHA}. Finally, in
Section~\ref{sec:dynamicMapping} we introduce the dynamic
reconfiguration mechanism, briefly discussing also how to render, in such
reconfigurations, partition strategies developed for hybrid
simulation algorithms.

\section{Transition-driven Stochastic Hybrid Automata}\label{sec:TDHA}

We define here a stochastic variant of \emph{Transition-Driven Hybrid
Automata}, introduced in~\cite{SB:Bortolussi:2009:HybridsCCPLattice}
as an intermediate layer to map \sCCP\ into hybrid automata. The
emphasis is on \emph{transitions} which, as always in hybrid automata, can be either discrete
(corresponding to jumps) or continuous (representing flows acting on
system's variables). The stochastic variant defined below contains two kind
of discrete transitions: instantaneous---as
in~\cite{SB:Bortolussi:2009:HybridsCCPLattice}---and stochastic,
which happen with an hazard given by a rate function.

\begin{definition}\label{def:TSHS}
A Transition-Driven Stochastic Hybrid Automaton (TDSHA) is a tuple\\
$\calT = (Q,\vr{X},\TC,\TD,\TS,\init{})$, where:
\begin{itemize}
\item $Q$ is a finite set of \emph{control modes}.

\item $\vr{X} = \{X_1,\ldots,X_n\}$ is a set of real valued
\emph{system's variables}\footnote{Notation: the time
  derivative of $X_j$ is denoted by $\dot{X_j}$, while the value
of $X_j$ after a change of mode is indicated by $X_j'$}.


\item $\TC$ is the set of \emph{continuous transitions or flows},
whose elements $\tau$ are triples $(q,stoich,rate)$, where:
$q\in Q$ is a mode, $stoich$ is a vector of size $|\vr{X}|$, and
$rate:\bbR^n\rightarrow \bbR$ is a (sufficiently smooth)
function. The elements of a triple $\tau$ are indicated by
$\cmode{\tau}$, $\stoich{\tau}$, and $\rate{\tau}$,
respectively.

\item $\TD$ is the set of \emph{instantaneous transitions}, whose elements $\delta$ are tuples of the
form\\
$(q_1,q_2,priority,guard,reset)$, where: $q_1$ is the
\emph{exit-mode}, $q_2$ is the \emph{enter-mode},
$priority:\bbR^n\rightarrow \bbR^+$ is a weight function used to
resolve non-determinism between two or more active transitions.
Moreover, $guard$ is  a quantifier-free first-order formula with
free variables in $\vr{X}$, representing the \emph{closed set}
$G_{\delta} = \{\vr{x}\in\bbR^n~|~guard[\vr{x}]\}$ in which thew
transition is active, and $reset$ is a deterministic update of
the form $\vr{X'} = f(\vr{X})$.\footnote{Even though there is no
real additional difficulty in considering stochastic
resets---i.e. in assuming $reset$ to be a transition
measure---we decided to avoid such move for the sake of
simplicity.}
%
%
The elements of a tuple $\delta$ are indicated by
$\exit{\delta}$, $\enter{\delta}$, $\priority{\delta}$,
$\guard{\delta}$, and $\reset{\delta}$, respectively.

\item $\TS$ is the set of \emph{stochastic transitions}, whose
elements $\eta$ are tuples of the form\\ $\eta =
(q_1,q_2,guard,reset,rate)$, where $q_1$, $q_2$, $guard$, and
$reset$ are as for transitions in $\TD$, while
$rate:\bbR^n\rightarrow \bbR^+$ is the rate function giving the
hazard of taking transition $\eta$. Such function is referred to
by $\rate{\eta}$.

\item $\init{}$ is a point giving the initial state of the system.
\end{itemize}
\end{definition}

A TDSHA has three types of transitions. Continuous transitions
represent flows and, for each $\tau \in \TC$, $\stoich{\tau}$ and
$\rate{\tau}$   give the \emph{magnitude} and the \emph{ form} of
the flow of $\tau$ on each variable $X\in\vr{X}$, respectively (see
below). Instantaneous transitions represent actions happening
immediately when their guard becomes true. Finally, stochastic
transitions happen at a specific rate. Both instantaneous and
stochastic transitions can change system variables according to a
specific reset function, depending on the variables'value at the
point in time at which the jump occurs.

\begin{remark}\label{rem:prioritiesVZrates}
Both priority and rates introduced in Definition~\ref{def:TSHS} make
TDSHA stochastic. Priorities define, at each point, a discrete
distribution of a random variable choosing among enabled
instantaneous transitions. Rates, on the other hand, define a random
race in continuous time, giving the delay for the next spontaneous
jump.
\end{remark}

\paragraph{Product of TDSHA.}
Given two TDSHA $\calT_1 =
(Q_1,\vr{X_1},\TC_1,\TD_1,\TS_1,\init{1})$ and\\ $\calT_2 =
(Q_2,\vr{X_2},\TC_2,\TD_2,\TS_2,\init{2})$, the product TDSHA $\calT
= \calT_1\otimes \calT_2$ can be defined in a simple way, along the
 path outlined in~\cite{SB:Bortolussi:2009:HybridsCCPLattice}. Essentially,
the discrete states'space  of the product automaton is $Q_1\times
Q_2$, while transitions from state $(q_1,q_2)$ are all those issuing
from $q_1$ \emph{or} $q_2$. Instantaneous or stochastic transitions
of $\calT_1$ going from state $q_1$ to state $q_1'$, will go from a
state $(q_1,q_2)$ to $(q_1',q_2)$ for each $q_2\in Q_2$.
Symmetrically for transitions of $\calT_2$.

\paragraph{Dynamics of TDSHA.} In order to formally define the
dynamical evolution of TDSHA, we can map them into a well-studied
model of Stochastic Hybrid Automata, namely Piecewise Deterministic
Markov Processes~\cite{STOC:Davis:1993:PDMP}. In this sense, TDSHA
are related to communicating
PDMP~\cite{HA:Strubbe:2007:CommunicatingPDP}, as they can also be
seen as a compositional formalism to model PDMP.
Due to space constraints, we just sketch here an informal
description of PDMP. The interested reader can find a more formal
treatment of PDMP and of their relation with TDSHA in the
supplementary material~\cite{supp}.

Basically, PDMP are stochastic processes whose \emph{state space} is
given by a finite collection of \emph{discrete modes} and by a set
of \emph{real-valued variables}. Within each mode, the continuous
variables evolve following the solution of a set of
\emph{mode-specific ODE's}. While in a mode, variables must stay
within the \emph{allowed region}. If they touch the boundary of the
allowed region, a \emph{forced discrete transition} is taken, and
the system may change mode and/or reset the value of the variables.
Moreover, the system is subject to the happening \emph{discrete
stochastic events}, governed by an hazard rate that is function of
the discrete mode and of continuous variables. Also stochastic
transitions trigger a reset of the state of the system.
\\
The main points of the mapping from TDSHA to PDMP are the following.

\begin{itemize}
  \item Within each discrete mode $q\in Q$, the system follows the
  solution of a system of ODE, constructed combining the effects
  of the continuous transitions $\tau$ acting on mode $q$.
  Essentially, the ODE for variable $X_i$ is obtained by adding
  up the rate of all such $\tau$ times the $i$-th component of
  the vector $\stoich{\tau}$:
  $$\dot{\vr{X}} = \sum_{\tau|~|\cmode{\tau} = q}
  \stoich{\tau}\rate{\tau}\ \ \ \mbox{in mode $q\in\Q$.}$$
  \item Two kinds of discrete jumps are possible: stochastic
  transitions are fired according to their rate, while
  instantaneous transitions are fired as soon as their guard
  becomes true. In both cases, the state of the system is reset
  according to the policy specified by \textbf{reset}. Choice
  among several active stochastic or instantaneous transitions
  is resolved probabilistically according to their rate or
  priority, see Remark~\ref{rem:prioritiesVZrates}.
  \item A trace of the system is therefore a sequence of (random) jumps
  interleaved by periods of continuous evolution.
\end{itemize}

\section{Stochastic Concurrent Constraint
Programming}\label{sec:sCCP}

In this section we briefly present (a simplified version of)
stochastic Concurrent Constraint Programming
(\sCCP~\cite{PA:Bortolussi:2006:sCCP}, a stochastic extension of
CCP~\cite{PA:Saraswat:1993:CCP}), as it seems to be sufficiently
expressive, compact, and especially easy to manipulate for our
purposes\footnote{There are other probabilistic extensions of CCP
studied in literature, like
\cite{PA:DiPierro:1998:pCCP,PA:Gupta:1997:PCCP,PA:Gupta:1999:PCCP}.
\cite{PA:DiPierro:1998:pCCP} provides CCP with a semantics based on
discrete time Markov Chains, while in
\cite{PA:Gupta:1997:PCCP,PA:Gupta:1999:PCCP} the stochastic
ingredient is introduced by extending the store with random
variables and adding a primitive for sampling. These approaches,
however, are not suited for our purposes, as we need a model in
which events happen probabilistically in continuous-time, as
customary in biochemical modeling.}. In the following we just sketch
the basic notions and the concepts needed in the rest of the paper.
More details on the language can be found
in~\cite{PA:Bortolussi:2006:sCCP,SB:Bortolussi:2008:BiomodelingSCCP:Journal}.

\begin{definition}\label{def:sCCP}
A \sCCP\ program is a tuple $\calA = (A,\calD,\vr{X},init(\vr{X}))$,
where
\begin{enumerate}
\item The \emph{initial network of agents} $A$
and the \emph{set of definitions} $\calD$ are given by  the
following grammar:
$$\begin{array}{c}
\calD =  \emptyset~|~\calD \cup\calD~|~\{C\defeq M\} \\
\pi =  [g(\vr{X}) \rightarrow u(\vr{X},\vr{X'})]_{\lambda(\vr{X})} \ \ \ \
M = \pi.C~|~M+M\ \ \ \
A =  M~|~A\parallel A
\end{array}$$
\item $\vr{X}$ is the set of variables of the store (with global
scope);
\item $init(\vr{X})$ is a predicate on $\vr{X}$ of the form $\vr{X}
= \vr{x_0}$, assigning an \emph{initial value} to store
variables.
\end{enumerate}
\end{definition}

In the previous definition, basic actions are \emph{guarded updates}
of (some of the) variables: $g(\vr{X})$ is a quantifier-free first
order formula whose atoms are inequality predicates on variables
$\vr{X}$ and $u(\vr{X},\vr{X'})$ is a predicate on $\vr{X},\vr{X'}$
of the form $\vr{X'} = f(\vr{X})$ ($\vr{X'}$ denotes variables of
$\vr{X}$ after the update), for some function
$f:\bbR^n\rightarrow\bbR^n$. Each such action has a \emph{stochastic
duration}, specified by associating  an exponentially distributed
random variable to actions, whose rate depends on the state of the
system through a function $\lambda:\vr{X}\rightarrow \bbR^+$.

\begin{example}\label{ex:main}
We will illustrate the notions introduced in the paper by means of
an example coming from biological systems. Specifically, we consider
a simple model of a (procaryotic) genetic regulatory network with a
single gene, expressing a protein acting, after dimerization, as a
repressor of its own production. We assume a cooperative repression:
two dimers are required to bind to the promoter region of the gene.
The \sCCP\ model is given by $\calA = (A,\calD,\vr{X},init)$, where
the variables are $\vr{X} = \{X_p,X_{p2}\}$, storing the quantity of
the protein $p$ and of its dimer $p2$ and the components in $\calD$
are (* stands for true):

\

{\tt
\begin{tabular}{lcl}
    gene$_0$ & $\defeq$ &  $[*\rightarrow X_p' = X_p + 1]_{k_{p1}}$.gene$_0$ +
                $[X_{p2} > 0 \rightarrow  *]_{k_{p1}X_{p2}}$.gene$_1$\\
    gene$_1$ & $\defeq$ &  $[*\rightarrow X_p' = X_p + 1]_{k_{p2}}$.gene$_1$ +
                $[X_{p2} > 0 \rightarrow  *]_{k_{p2}X_{p2}}$.gene$_2$ + \\
                && $[*\rightarrow *]_{k_{u1}}$.gene$_0$\\
   gene$_2$ & $\defeq$ &  $[*\rightarrow *]_{k_{u2}}$.gene$_1$\\
    deg & $\defeq$ &  $[* \rightarrow X_p' = X_p - 1]_{k_d X_p}$.deg\\
   dimer & $\defeq$ & $[* \rightarrow X_p' = X_p - 2 \wedge X_{p2}' = X_{p2} + 1]_{k_x  X_p ( X_p - 1 ) / 2}$.dimer +\\
           && $[* \rightarrow X_p' = X_p + 2 \wedge X_{p2}' = X_{p2} - 1]_{k_{-x} X_{p2}}$.dimer\\
\end{tabular}

}

\

The initial network $A$ is
    {\tt gene$_0 \parallel$ deg $\parallel$ dimer}
with initial values of the store variables are given by
$$init(X_p,X_{p2}) = (X_p = 0)\wedge (X_{p2} = 0).$$

Notice: there is no need to introduce agents for proteins or dimers,
as the quantity of these objects needs only to be measured by stream
variables. The repression mechanism is represented by a gene unable
of expressing a protein whenever in state {\tt gene$_2$}. We did not
decrement $X_{p_2}$ before entering states {\tt gene$_1$} and {\tt
gene$_2$} as we assume repression mechanism not requiring a binding
of the dimer (inhibition by bumping).
\end{example}

\begin{remark}\label{rem:bioSCCP}
The pros and cons of using \sCCP\ as a modeling language for
biological systems are discussed in detail
in~\cite{SB:Bortolussi:2008:BiomodelingSCCP:Journal}. Basically,
\sCCP\ combines on one side the logical simplicity of process
algebras and on the other side the computational power of
constraints. As a matter of fact, the constraint store can be more
general than that used in this paper, whereby more complex information
(like spatiality) can be managed just by a simple programming
activity. Further work is needed, however, to export the techniques
developed here to a more general version of the store.
\end{remark}

All agents definable in \sCCP, i.e. all agents $C\defeq M \in
\calD$,\footnote{In the following, with a slight abuse of
notation, we sometimes write $C\in\calD$ for $C\defeq M \in
\calD$.} are \emph{sequential}, i.e. they do not contain any
occurrence of the parallel operator, whose usage is restricted
at the upper level of the network.
\\
\sCCP\ sequential agents can be seen as automata synchronizing
on store variables and they can be conveniently represented as
labeled graphs, called \emph{Reduced Transition Systems} (RTS)
(see~\cite{SB:Bortolussi:2007:QAPL:SPAandODE}).

\begin{figure}[!h]
  \begin{center}
  \includegraphics[height=4cm]{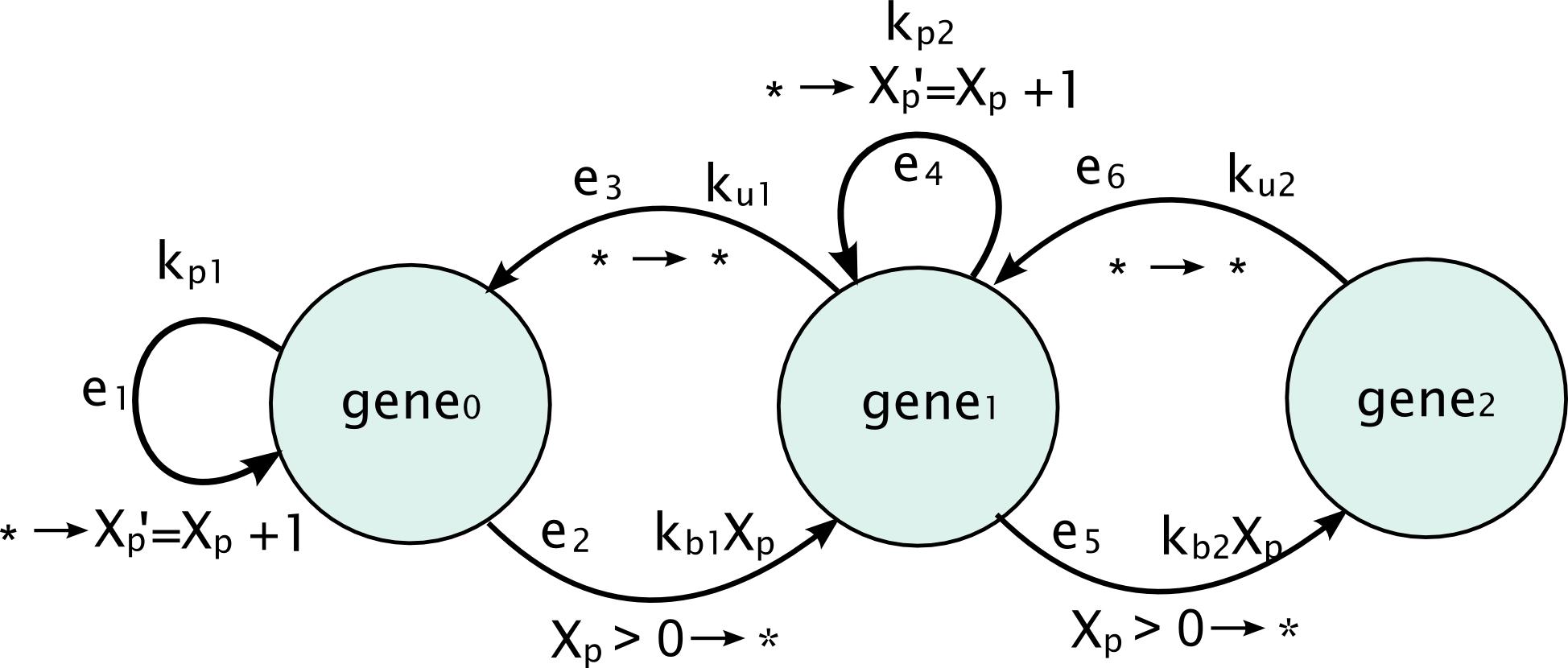}
  \end{center}
  \caption{Reduced Transition Systems for the agent {\tt gene$_0$} defined in Example~\ref{ex:main}.
  Each edge is labeled by its rate function
and by the guard and the update (depicted in the $guard \rightarrow update$ notation).}\label{fig:RTS_example}
\end{figure}

The steps to obtain
an object suitable to our subsequent treatment are the following:
\begin{enumerate}
  \item Define the collection of all possible states---the
\emph{derivative set} $Der(C)$---and actions---$action(C)$---of
any sequential agent appearing in a \sCCP\ program.

\item
Restrict to \sCCP\ \emph{simple programs}, i.e. programs without
multiple copies of the same agent running in parallel at the
same time. Formally, it is required that the derivative sets of
any two agents in parallel in the initial network are disjoint.
This is only an apparent restriction,
cf.~\cite{SB:Bortolussi:2009:HybridsCCPLattice} for a more
detailed discussion.

  \item Introduce the following multi-graph\footnote{$exit(\pi),enter(\pi),guard(\pi),update(\pi),rate(\pi)$
  give the executing agent, the target agent, the
  guard, the update and the rate of an action $\pi$, respectively.}  $\text{\emph RTS}(C) = (S(C),E(C),\ell)$:
\begin{itemize}
  \item $S(C) = Der(C)$,
  \item $E(C) = \{(exit(\pi),enter(\pi))~|~\pi\in action(C)\}$,
  \item $\ell(e) = (guard(\pi),update(\pi),rate(\pi))$, where $\pi$
  is the action defining  $e\in E(C)$.
\end{itemize}
In Figure~\ref{fig:RTS_example}, we show the RTS for the agent
{\tt gene$_0$}, defined in Example~\ref{ex:main}.
    \item Introduce the notion  of \emph{extended} \sCCP\ program
    $$\calA^+ = \sccptuple{X \mbox{$\cup \{P_C~|~C \in \cal D\}$} }{}{+},$$ in which a variable $P_C$ for
    run-time recording  the number of parallel copies of each
    agent $C \in \calD$ is available, and prove $\calA^+$ is isomorphic to
    $\cal A$ (see~\cite{SB:Bortolussi:2009:HybridsCCPLattice}
    for further details).
\end{enumerate}
Essentially, the last step is a technical trick that simplifies the
overall treatment. The variable $P_C$ counts the
number of copies of $C$ present in parallel within the system at a
given point in time. To take into account the effects of transitions
on agents, we modify updates and rate functions, by
increasing/decreasing counter $P_C$ relative to actions
adding/removing a copy of $C$. The reason for introducing state
variables will be apparent in next section. They are required to
control a cluster of discrete states (continuously approximated) and
the \emph{real} value of a state variable will indicate the
``tendency'' of the system to be in that particular state.

\section{From sCCP to TDSHA}\label{sec:sCCPtoTDHA}

In this section we define a semantics for \sCCP\ in terms of
 TDSHA. The basic idea is to partition all possible
transitions executable by an \sCCP\ agent into two classes: those
remaining discrete-stochastic and those to be approximated as
continuous. Different partitions'schemata correspond to different
TDSHA. By parameterizing upon such schemata, we will obtain a
\emph{lattice} of different TDSHA's.
\\
Note that this approach is different
from~\cite{SB:Bortolussi:2009:HybridsCCPLattice}, as \emph{we do not
remove stochasticity}. Indeed, Stochastic Hybrid Automata can be
seen as an intermediate layer between stochastic programs and
(non-stochastic) hybrid systems. The reader is referred
to~\cite{SB:Bortolussi:2009:HybridSCCPstaticJournal} for further
discussions.

The mapping proceeds in two steps. First we convert into TDSHA's
each sequential component of a \sCCP\ program, then all these
TDSHA's are combined using the product construction.

Given a $\calA^+ = \sccptuple{Y}{}{+}$,  let $C\in \calD^+$ be one
of the components of the initial network $A$, and let $RTS(C) =
(S(C),E(C),\ell)$ be its RTS.

A specific continuous/discrete scheme of approximation is formalized
by the choice of a boolean vector $\kappa\in\{0,1\}^m$, $m=|E(C)|$,
indexed by edges in $E(C)$: for $e\in E(C)$, $\kappa[e] = 1$ stands
for a continuous approximation of the transition, while $\kappa[e] =
0$ implies that the transition will remain discrete. Let
$E(\kappa,C)= \{e\in E(C)|~\kappa[e] = 1\}$  and
 $E(\neg\kappa,C)= \{e\in E(C)|~\kappa[e] = 0\}$.

In order to guarantee that the vector field constructed from
continuous transitions is sufficiently regular, we identify as
\emph{continuously approximable} only those actions $\pi$ such that
$rate(\pi)$ is differentiable and $rate(\pi)[\vr{X}] = 0$ whenever
$guard(\pi)[\vr{X}]$ is false.\footnote{Guards of continuosly
approximable $\pi$ are, in fact, redundant.} We call
\emph{consistent} a vector $\kappa$ such that $\kappa[e] = 1$ only
for edges $e$ that are continuously approximable. In the following,
we suppose to work only with consistent $\kappa$.

At this point we are ready to introduce the basic components of our target TDSHA.

\paragraph{Discrete Modes.}
The modes of the TDSHA will be essentially the states $S(C)$ of the
$RTS(C)$. However, as continuous transitions cannot change mode, we
need to consider as equivalent those states that can be reached by a
path of continuous edges.
Let us denote by $\sim_\kappa$ the equivalence relation among states
of $S(C)$ relating two states if and only if they are connected by a path
of continuous edges (i.e. edges in $E(\kappa,C)$ of the non-oriented
version of $RTS(C)$). Let $\mode{\kappa}{C} = S(C)/\sim_\kappa$. For
each edge $e\in E(\kappa,C)$, we define the \emph{stoichiometric
vector} $\nu_{\vr{Y},e}$ as an $|\vr{Y}|$-vector,   $\vr{Y} =\vr{X} \cup \{P_C~|~C \in \cal D\}$, such that
$\nu_{\vr{Y},e}[X]=h$ if and only if variable $X$ is updated by
transition $e$ according to the formula $X' = X + h$.

\paragraph{Example.} Consider the gene component of
Example~\ref{ex:main}. Its RTS, shown in
Figure~\ref{fig:RTS_example}, has three states, corresponding to the
three components $\mathrm{gene}_0$, $\mathrm{gene}_1$, and
$\mathrm{gene}_2$, with state variables denoted by $P_0$, $P_1$, and
  $P_2$, respectively. The RTS has also 6 transitions, indexed by
$e_1,\ldots,e_6$. Consider the $\kappa$ vector equal to
$(1,0,0,1,1,1)$: edges $e_1,e_4,e_5,e_6$ will be approximated as
continuous, while the other three remain discrete. The relation
$\sim_\kappa$ has a quotient state space containing two classes:
$S_1 = \{\mathrm{gene}_0\}$ and $S_2 =
\{\mathrm{gene}_1,\mathrm{gene}_2\}$. Such a partitioning of the
gene's states can be seen  as a way to render a slower dynamics for
the binding/unbinding mechanism of the first repressor, to be
compared to a faster one relative the second copy of the repressor.

\paragraph{Continuous flow.}

The continuous evolution for TDSHA is given by the following set of
continuous transitions:
$$\TC =
\{([exit(e)],\nu_{\vr{Y},e},rate(e)~|~e\in
E(\kappa,C)\}.$$

\paragraph{Stochastic transitions.}

Stochastic transitions are defined in a very simple way, as guards
and rates are basically copied from the $\sCCP$ edge. The only
technicality  is the definition of the reset.

Consider the state counting variables $\vr{P}=\{P_C~|~C \in \cal
D\}$. They can assume values less than or equal to one, as the
initial program is \emph{simple}. Moreover, they range in the whole
real-valued interval $[0,1]$ whenever we are  in a clustered state
$[s]$ collapsing $\xs{s}{1}{k}$ of $\text{\emph{ RTS}(C)}$. In this
case, the state variables $P_{s_1},\ldots,P_{s_k}$  must sum exactly
to 1, their value representing the likelihood of state $s_1, \ldots
, s_k$ of the cluster $[s]$, respectively. In order to deal with
state clusters correctly, we have to ensure that when a state $[s]$
is left, all its state variables are set to zero. Moreover, if a
discrete transition looping in $[s]$ takes place, then the variable
of its target state $s_i$ must be set to 1, while all other
variables of $[s]$ are to be reset to 0. To enforce this, consider
an $\sCCP$ edge connecting states $s_1$ and $s_2$, with
$$update(e) \defeq \vr{X'} = f(\vr{X}) \wedge P_{s_1}' = P_{s_1} - 1
\wedge P_{s_2}' = P_{s_2} + 1,$$ and define the function $f_P$ on
$\vr{P}$ which is 1 on the component corresponding to $P_{s_2}$ and
zero elsewhere. In this way, $\vr{P'} = f_P(\vr{P})$ implements the
correct updating policy. Let now $\bar{f}$
combine $f$ and $f_P$: $\bar{f}\begin{pmatrix}  \vr{X} \\
\vr{P} \\ \end{pmatrix} = \begin{pmatrix} f(\vr{X}) \\ f_P(\vr{P}) \\
\end{pmatrix}$.

Putting everything together, we have that the discrete transition
associated with $e\in E(\neg\kappa,C)$ with $e=(s_1,s_2)$ is
$$\begin{array}{c}([s_1],[s_2],guard(e),\vr{Y'} = \bar{f}(\vr{Y}),rate(e))\in \TS.\end{array}$$

\paragraph{Instantaneous transitions.} At this stage, there is no
need to define instantaneous transitions. They will be used in
Section~\ref{sec:dynamicMapping} to deal with dynamic partitioning.

\

We can now collect all our considerations into the following definition.

\begin{definition}\label{def:TDHAsequentialAgent}
Let $\calA = (A,\calD,\vr{X},init_0)$ be a simple \sCCP\ program and
$\calA^+ = (A^+,\calD^+,\vr{Y},init_0^+)$ be its extended version.
Let $C$ be a sequential component in parallel in $A^+$, with $RTS(C)
= (S(C),E(C),\ell)$. Fix a boolean vector $\kappa\in\{0,1\}^m$, $m =
|E(C)|$. The Transition-Driven Stochastic Hybrid Automaton
associated with $C$ with respect to  $\kappa$ is $\TDHA{}(C,\kappa)
= (Q,\vr{Y},\TC,\TD,\TS,\init{})$, where
\begin{itemize}
  \item $Q = \mode{\kappa}{C} = S(C)/\sim_\kappa$;
  \item $\TC = \{([exit(e)],\nu_{\vr{Y},e},rate(e)~|~e\in E(\kappa,C)\}$;
\item $\TD = \emptyset$;
  \item $\TS = \{([s_1],[s_2],guard(e),\vr{Y'} = \bar{f}(\vr{Y}),rate(e))~|~e=(s_1,s_2)\in E(\neg\kappa,C)\}$;
  \item $\init{} = init_0$.
\end{itemize}
\end{definition}

\begin{figure}[!t]
  \begin{center}
  \includegraphics[height=4cm]{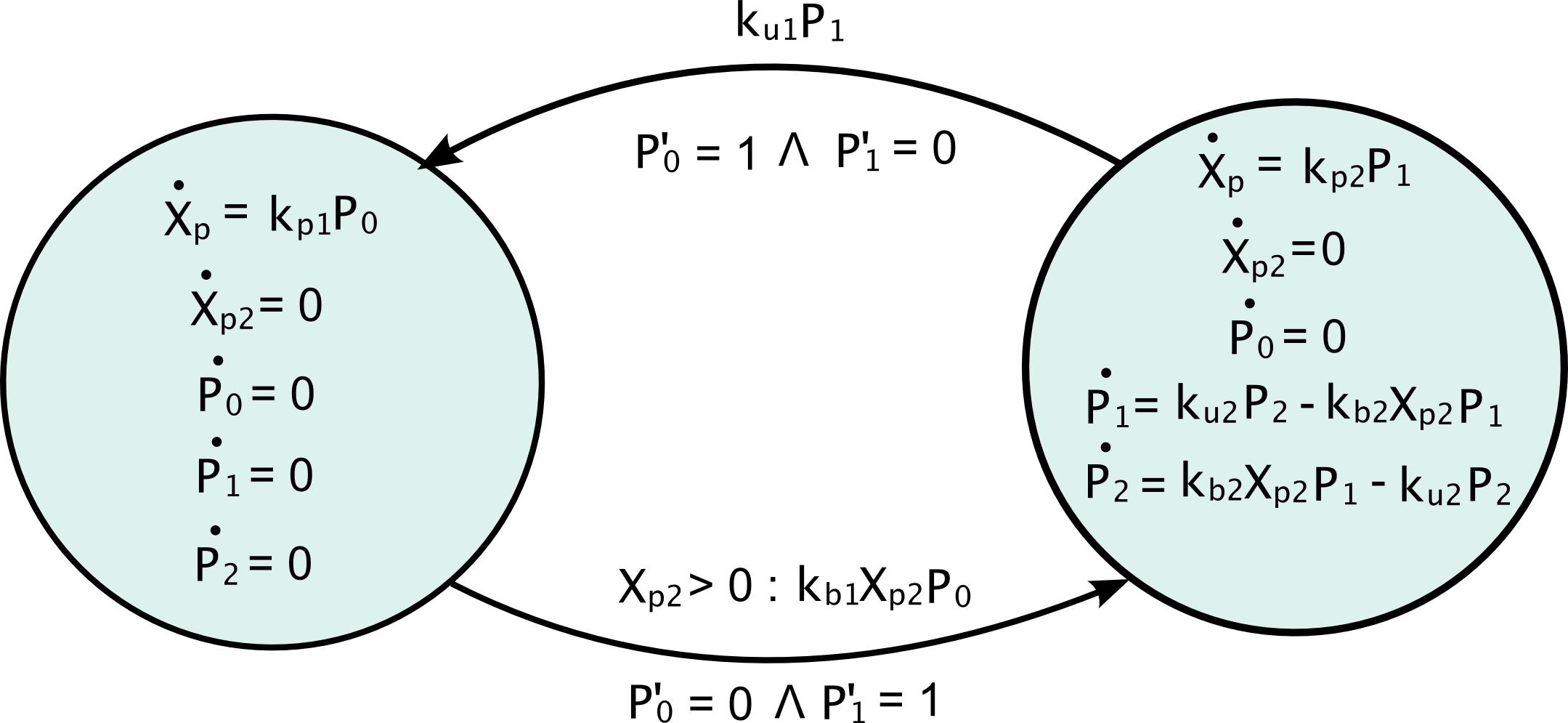}
  \end{center}
  \caption{PDMP associated with the gene component of Example~\ref{ex:main}. Transitions approximated continuously
  determine a set of ODE, while discrete transitions are stochastic and are depicted here as edges of a graph
  (showing rates, guards and resets). The ODEs are obtained from continuous transitions according to
  the recipe of Section~\ref{sec:TDHA}. We chose to display TDSHA in this format as it is similar
  to the classic representation of hybrid automata.}\label{fig:HAgeneComp}
\end{figure}

\paragraph{Example.} From the previous definition it is easy to generate
the TDSHA relative to our running example above, in which $\kappa =
(1,0,0,1,1,1)$. Once we have the TDSHA, we can generate the
corresponding PDMP (see supplementary material~\cite{supp}), which
is shown in Figure~\ref{fig:HAgeneComp}.
\bigskip

Definition \ref{def:TDHAsequentialAgent} gives a way to associate a
TDSHA with a sequential agent of a \sCCP\ program. In order to
define the TDSHA for the whole program, we will use the product
construction.

\begin{definition}\label{def:TDHAsCCPprogram}
Let $\calA = (A,\calD,\vr{X},init_0)$ be a simple \sCCP\ program and
$\calA^+ = (A^+,\calD^+,\vr{Y},init_0^+)$ be its extended version,
with $A^+ = C_1\parallel\ldots\parallel C_n$. Fix a boolean vector
$\kappa_i$ for each sequential agent $C_i$. The Transition-Driven
Hybrid Automaton for the \sCCP\ program $\calA$, with respect to
$\kappa=(\kappa_i)_{i=1,\ldots n}$ is
$$\TDHA{}(A,\kappa) =
\TDHA{}(C_1,\kappa_1)\otimes\cdots\otimes\TDHA{}(C_n,\kappa_n).$$
\end{definition}

\begin{figure}[!t]
  \begin{center}
  \includegraphics[height=5cm]{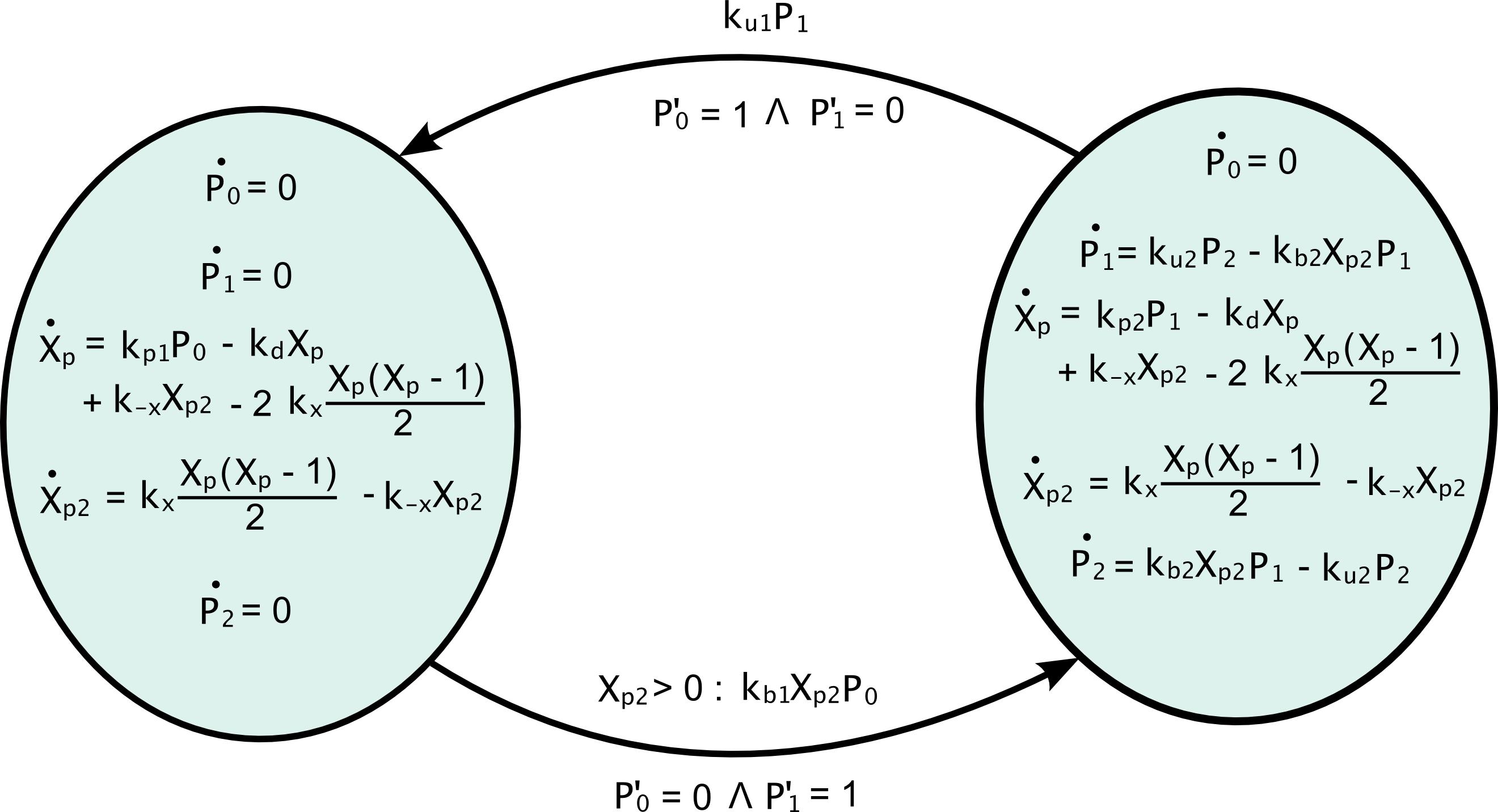}
  \end{center}
  \caption{PDMP obtained from the product of TDSHA associated with the three components
   of the \sCCP\ program of Example~\ref{ex:main}. See also the caption of Figure~\ref{fig:HAgeneComp}.}\label{fig:HAexample}
\end{figure}

\paragraph{Example.}
Consider again the \sCCP\ model of Example~\ref{ex:main}. It has
three components: gene, deg and dimer, with 6, 1, and 2 edges
respectively. We consider three vectors $\kappa_1 = (1,0,0,1,1,2)$,
$\kappa_2 = (1)$, and $\kappa_3 = (1,1)$. The product TDSHA of these
three components generates the PDMP depicted in
Figure~\ref{fig:HAexample}.

\subsection{Lattice of TDSHA}\label{sec:latticeTDHA}

Definition~\ref{def:TDHAsCCPprogram} associates a TDSHA with a
\sCCP\ agent for a fixed partition, given by  vector $\kappa$, of
the transitions into discrete and continuous. Clearly, different
choices of $\kappa$ correspond to different TDSHA's, with a
different degree of approximation of the original \sCCP\ program.
The different TDSHA's  can be arranged into a lattice according to
the following pre-order:
\begin{definition}\label{def:ordering}
Let $\calA$ be a \sCCP\ agent, then $\TDHA{}(\calA,\kappa_1)
\sqsubseteq \TDHA{}(\calA,\kappa_2)$ if and only if $\kappa_1[e] = 1
\Rightarrow \kappa_2[e] = 1$, for each transition $e\in
E(A)=E(C_1)\cup\ldots\cup E(C_n)$, with
$A=C_1\parallel\ldots\parallel C_n$ the initial agent of $\calA$.
\end{definition}

The bottom element of this lattice is obtained for $\kappa\equiv 0$,
while the top element is obtained for $\kappa[e] = 1$ if and only if
$e$ is continuously approximable. We remind to the reader that
transitions not continuously approximable \emph{must} be kept
discrete.

The two ``extreme'' choices correspond to two particularly important TDSHA's, as shown in the
following propositions.

\begin{proposition}\label{prop:bottomElement}
Let $\calA$ be a \sCCP\ program. The TDSHA $\TDHA{}(\calA,0)$ is the
CTMC associated with $\calA$ by its standard semantics.
\end{proposition}

\begin{proposition}\label{prop:topElement}
Let $\calA$ be a \sCCP\ program with initial agent
$A=C_1\parallel\ldots\parallel C_n$. If $e$ is continuously
approximable for each $e\in E(A) = E(C_1)\cup\ldots\cup E(C_n)$,
$\TDHA{}(\calA,\vr{1})$ coincides with the system of ODE's
associated with $\calA$ by its fluid-flow approximation
(see~\cite{SB:Bortolussi:2007:QAPL:SPAandODE}).
\end{proposition}

\section{Dynamic Partitioning of
Transitions}\label{sec:dynamicMapping}

In the previous sections we have defined a mapping from \sCCP\ to
TDSHA fixing the level of discreteness and continuity. This choice,
however, can be difficult to perform \emph{a priori}, as one does not know
if the system will evolve to a state where a different approximation
is more accurate.
\\
This is particularly true when one deals with biological systems. In
this case, reactions involving large populations of molecular
species or having high rates, may be treated as continuous. However,
such conditions depend on the state of the system and may change
during its evolution. Indeed, there has been a growing attention on
hybrid simulation strategies in systems biology, cf. next Section.

In order to have dynamic switching, we can extend the discrete
modes, introducing states for each admissible vector $\kappa$. New
discrete transitions need to be added as well, changing the value of
$\kappa$ according to some user defined conditions.
\\
Intuitively, for each transition $e$ the state space must be
partitioned in two---possibly empty---regions: one where the edge
$e$ is treated as a continuous transition and one in which it is
discrete. In order to define such regions, we consider two
predicates:
\begin{enumerate}
  \item $cont[e](\vr{X})$, encoding the condition to change edge $e$ from discrete to
  continuous;
  \item $disc[e](\vr{X})$, encoding the condition to change edge $e$ from continuous to
  discrete.
\end{enumerate}
An obvious requirement is that the regions identified by $cont[e]$
and $disc[e]$ should be  disjoint. In order to define such predicates, we
will consider a (sufficiently regular, usually continuous) function
$f_e:\bbR^n\rightarrow\bbR$, whose sign will discriminate between
continuous and discrete regions for edge $e$. More specifically, we
define, for a fixed, small $\eps>0$.
\begin{enumerate}
  \item $cont[e](\vr{x}) := f_e(\vr{x})\geq \eps$;
  \item $disc[e](\vr{x}) := f_e(\vr{x})\leq -\eps$.
\end{enumerate}
Using $\eps$ instead of 0, not only guarantees that the regions in
which $cont[e]$ and $disc[e]$ are true are disjoint, but also avoids
pathological situations of infinite sequences of instantaneous
transitions. See supplementary material~\cite{supp} for further
details.

%

Now, suppose $e$ is continuous. If the current trajectory enters in
a region of the state space in which $disc[e]$ becomes true, then we
must trigger an instantaneous transition in order to move from
$\kappa_1[e]=1$ to $\kappa_2[e]=0$. All the variables must remain
unchanged. However, in this case it may happen that the new relation
$\sim_{\kappa_2}$ splits in two the current mode $[s]\in
S_{\kappa_1}(C)$, say $[s]_{\kappa_1} =
[s_1]_{\kappa_2}\cup[s_2]_{\kappa_2}$. In this case, we need to
introduce two instantaneous transitions, one going to $[s_1]$ and
the other to $[s_2]$.
Now, consider the value of the state variables of $[s]$, $P_{[s]} =
\sum_{s_i\in [s]} P_{s_i}$. It can be proved that $P_{[s]} = 1$.
Moreover, $P_{[s]} = P_{[s_1]} + P_{[s_2]}$ but, clearly, it is not
necessarily the case that the two quantities on the right hand side
of the equality are equal. This means that the system may ``prefer''
to move to states in $[s_1]$ than to those in $[s_2]$. This
situation is correctly modeled using priorities, i.e. weighting
transition to $[s_i]$ by $P_{[s_i]}$ and re-normalizing variables in
$[s_1]$ and $[s_2]$ to maintain the property $P_{[s]} = 1$ for each
$[s] \in S_{\kappa_2}(C)$.

We now give a formal definition for this construction, following a
similar strategy as in Section~\ref{sec:sCCPtoTDHA}: first we
construct TDSHA for sequential components, then we apply the product
construction to combine the local constructions. In order to fix the
notation, consider the TDSHA $\TDHA{}(C,\kappa) =
(Q,\vr{Y},\TC,\TD,\TS,\init{})$ associated with a component $C$,
with respect to a  fixed $\kappa$. With $Q_{\kappa}$ we indicate the
set $Q_{\kappa} = \{([s]_\kappa)~|~[s]\in Q\}$. Moreover,
$\TC_{\kappa}$, $\TD_{\kappa}$, and $\TS_{\kappa}$ denote the sets
$\TC$, $\TD$, and $\TS$, respectively, with states in $Q$ replaced
by the corresponding states (equivalence classes) in $Q_{\kappa}$. A
similar rule applies to $\init{\kappa}$.

\begin{definition}\label{def:dynamicTDHAcomponent}
Let $\calA = (A,\calD,\vr{X},init_0)$ be a simple \sCCP\ program and
$\calA^+ = (A^+,\calD^+,\vr{Y},init_0^+)$ be its extended version.
Let $C$ be a sequential agent in parallel in $A^+$, with $RTS(C) =
(S(C),E(C),\ell)$ and $|E(C)| = m$. Moreover, let $cont[e],disc[e]$,
$e\in E(C)$ be defined as above. The TDSHA with dynamic partitioning
associated with $C$ is $\TDHA{}(C,cont,disc) =
(Q,\vr{Y},\TC,\TD,\TS,\init{})$, with:

\begin{enumerate}
\item $Q = \bigcup_{\kappa\in\{0,1\}^m} Q_{\kappa}$;
\item $\TC = \bigcup_{\kappa\in\{0,1\}^m}\TC_{\kappa}$;
\item $\TS = \bigcup_{\kappa\in\{0,1\}^m}\TS_{\kappa}$;
\item $\TD = \bigcup_{\kappa\in\{0,1\}^m}\TD_{\kappa} \cup \TD_{0,1}
\cup \TD_{1,0}$, where
\begin{eqnarray*}
\TD_{1,0} & = & \Bigg\{\bigg([s_1]_{\kappa_1},
[s_2]_{\kappa_2}, P_{[s_2]}, disc[e], \vr{Y'} = g(\vr{Y})\bigg)~\bigg|~ e\in E(C),\\
& & \kappa_1(e) = 1,\kappa_2(e) = 0,
\kappa_1(e') = \kappa_2(e')\ for\ e\neq e', [s_1]_{\kappa_1}\cap [s_2]_{\kappa_2}\neq \emptyset \Bigg\}
\end{eqnarray*}
where $g$ assigns value $\frac{P_{s'}}{P_{[s]}}$ for $s'\in
[s_2]_{\kappa_2}$, 0 to any other $P_s$, and it is the identity
on $\vr{X}$. Moreover
\begin{eqnarray*}
\TD_{0,1} & = & \Bigg\{\bigg([s]_{\kappa_1},
[s]_{\kappa_2}, 1, cont(e), \vr{Y}' = \vr{Y}\bigg)~\bigg|~ e\in E(C),\\
& & \kappa_1(e) = 0,\kappa_2(e) = 1,
\kappa_1(e') = \kappa_2(e')\ for\ e\neq e' \Bigg\};
\end{eqnarray*}
\item $\init{} = init_0^+$;
\end{enumerate}
\end{definition}

\begin{definition}\label{def:dynamicTDHAsCCPprogram}
Let $\calA = (A,\calD,\vr{X},init_0)$ be a simple \sCCP\ program and
$\calA^+ = (A^+,\calD^+,\vr{Y},init_0^+)$ be its extended version,
with $A^+ = C_1\parallel\ldots\parallel C_n$. Moreover, fix
predicates $cont_j[e],disc_j[e]$ for each sequential agent $C_j$ of
$A^+$, according to Definition~\ref{def:dynamicTDHAcomponent}. The
Transition-Driven Stochastic Hybrid Automata with dynamic
partitioning for the \sCCP\ program $\calA$, with respect to
$(cont_j,disc_j)_{j=1,\ldots n}$ is
$$\TDHA{}(A,(cont_j,disc_j)_{j=1,\ldots n}) =
\TDHA{}(C_1,cont_1,disc_1)\otimes\cdots\otimes\TDHA{}(C_n,cont_n,disc_n).$$
\end{definition}

\begin{remark}[On the fly simulation]\label{rem:OnTheFlySim}
In Definition~\ref{def:dynamicTDHAsCCPprogram}, the resulting TDHA
has a number of modes exponential in the number of transitions that
sequential agents can perform. This combinatorial explosion rules
out the possibility of generating all the modes together. However,
if we restrict to simulation, this is not a real issue, as we need
to record only the current mode: the target mode of a transition can
be generated on the fly as soon as the transition has been taken,
given the knowledge of $RTS$.
\end{remark}

\subsection{Hybrid Simulation
Strategies}\label{sec:hybridPartiniongSB}

The hybrid simulation algorithms proposed in
literature~\cite{SB:Rawlings:2002:HybridSimulation,SB:Neogi:2004:HybridSimulationDynamicPartitioning,SB:HaseltineRawlings:2005:OriginsApproximationChemKin,SB:Kaznessis:2005:HybridSimulation,SB:Kiehl:2004:hybridSimulation,SB:Griffith:2006:HybridSimulation,SB:Alfonsi:2005:HybridSimulation}
basically differ in two aspects: the kind of continuous dynamics (it
can be based on ODE or SDE) and the rules for partitioning reactions
into continuous and discrete (usually called fast and slow). More
specifically, the partitioning can be static (done at the beginning
of the simulation) or dynamic (i.e. recomputed at run-time).

Conditions for separating fast and slow reactions are usually
twofold:
\begin{enumerate}
  \item the \emph{size of species} involved in the reaction must all be
  bigger than a given threshold. Usually, a fast reaction $j$
  must satisfy a condition like $x_i \geq K |\nu_{i,j}|$ for all
  species $i$ involved in $j$, where $\nu$ is the stoichiometric
  matrix.
  \item the \emph{rate function} of fast reactions must be reasonably
  bigger than that of slow reactions. Usually, the following
  constraint is
  enforced~\cite{SB:Kaznessis:2005:HybridSimulation}:
  $\lambda_j(\vr{x})\Delta t \geq \Lambda$, which ensures that
  reaction $j$ fires many times during the time step $\Delta t$.
  In~\cite{SB:Griffith:2006:HybridSimulation} a different
  partition strategy imposes that rates of fast reactions are
  $\Lambda$ times faster than the fastest slow reaction, so as
  to guarantee a separation of time scales.
\end{enumerate}

Dynamical policies sketched above can be easily accounted for in our
setting.
\\
First of all, we need to start from an \sCCP\ model of a biochemical
network~\cite{SB:Bortolussi:2008:BiomodelingSCCP:Journal}, in which
reactions are modeled by action capabilities of agents. Then,
applying the framework of this paper, we associate a TDSHA with such
a model, together with a suitable policy for dynamic partitioning of
transitions. All we have to do is define a function $f_e$ for each
\sCCP\ transition $e$, such that $f_e(\vr{x}) > 0$ when the
associated reaction can be considered fast and $f_e(\vr{x}) < 0$
when it is slow.

As an example, consider a partition strategy based only on the size
of populations, like the one adopted
in~\cite{SB:Neogi:2004:HybridSimulationDynamicPartitioning}. In this
case, the function $f_e$ for transition $e$ can be the following:

$$f_e(\vr{x}) = \min\{x_i-K|\nu[x_i,e]|~|~\nu[x_i,e]\neq 0\},$$
where $\nu[\cdot,e]$ is the stoichiometry of action $e$, constructed
as in Section~\ref{sec:sCCPtoTDHA}, and $K$ is a constant (that can
be tuned for the specific system). Of course, more complex policies
can be introduced by suitably modifying the functions $f_e$.

\section{Conclusion and Further Directions}\label{sec:conclusions}

In this paper we provided a specific process algebra, \sCCP, with a
general semantics based on stochastic hybrid systems, parametric
with respect to  the degree of continuity and discreteness. The
different hybrid models generated in this way can be arranged in a
lattice, and we provided also a way to dynamically move within the
lattice. This allows to formally describe hybrid simulation
algorithms, opening up their use as tools to simulate process
algebra-based models. Moreover, this approach gives the possibility
of using other computational analysis methods than simulation, like
reachability computations or model checking.
\\
An interesting problem is how to extend such machinery to other
process algebras. First steps have been done to deal with stochastic
$\pi$-calculus~\cite{PA:Bortolussi:2009:HybridPIJournal}, however
the peculiarities of each language present specific difficulties to
be solved.

The formal treatment developed in the paper, in particular the
lattice of TDSHA  defined in Section~\ref{sec:latticeTDHA}, can
also provide an interesting theoretical framework to study the
quality of the approximation and the error introduced. In
particular,  the mature theory of PDMP~\cite{STOC:Davis:1993:PDMP}
can provide interesting tools in this direction.

Another issue we are investigating regards the relationships between
discreteness and stochasticity. In particular, we are interested in
understanding whether the stochastic ingredient of the dynamics can
be dropped in favor of a pure discrete evolution, and at what
price~\cite{SB:Bortolussi:2009:HybridSCCPstaticJournal}. Motivations
for this reside in the fact that non-stochastic hybrid systems have
a much wider and more efficient set of automated reasoning tools available.

We conclude with a more basic (perhaps philosophical) question:
given that a mix of continuous and discrete simulation strategy is
the choice, is there a way---other than minimization of
computational complexity---to determine which parts of the systems
can/may be simulated discretely/continuously? We feel that  physical
consideration must be taken into account for addressing this issue
and that these are probably outside our reach. However we wish to
contribute the ``computer scientist point of view": the level of
discreteness/continuity can be established on the ground of a formal
specification of the properties to verify/simulate and should
guarantee the minimum of computational resources necessary to this
task.

\bibliographystyle{eptcs}

\begin{thebibliography}{10}
\providecommand{\bibitemstart}[1]{\bibitem{#1}}
\providecommand{\bibitemend}{} \providecommand{\bibliographystart}{}
\providecommand{\bibliographyend}{}
\providecommand{\url}[1]{\texttt{#1}}
\providecommand{\urlprefix}{Available at }
\providecommand{\bibinfo}[2]{#2}
\bibliographystart

\bibitemstart{supp}
\bibinfo{title}{Supplementary matherial to the paper available
online at: } \url{http://www.dmi.units.it/~bortolu/sccp.htm}.
\bibitemend


\bibitemstart{SB:Alfonsi:2005:HybridSimulation}
\bibinfo{author}{A.~Alfonsi}, \bibinfo{author}{E.~Cances},
  \bibinfo{author}{G.~Turinici}, \bibinfo{author}{B.~Di Ventura} \&
  \bibinfo{author}{W.~Huisinga} (\bibinfo{year}{2005}):
  \emph{\bibinfo{title}{Adaptive simulation of hybrid stochastic and
  deterministic models for biochemical systems}}.
\newblock In: {\sl \bibinfo{booktitle}{Proceedings of ESAIM}},
  ~\bibinfo{volume}{14}. pp. \bibinfo{pages}{1--13}.
\bibitemend

\bibitemstart{PA:Bortolussi:2006:sCCP}
\bibinfo{author}{L.~Bortolussi} (\bibinfo{year}{2006}):
  \emph{\bibinfo{title}{Stochastic Concurrent Constraint Programming}}.
\newblock In: {\sl \bibinfo{booktitle}{Proceedings of 4th International
  Workshop on Quantitative Aspects of Programming Languages (QAPL 2006)}}, {\sl
  \bibinfo{series}{ENTCS}} \bibinfo{volume}{164}. pp. \bibinfo{pages}{65--80}.
\bibitemend

\bibitemstart{PA:Bortolussi:2007:SCCPandODEjournal}
\bibinfo{author}{L.~Bortolussi} \& \bibinfo{author}{A.~Policriti}
  (\bibinfo{year}{2009}): \emph{\bibinfo{title}{Dynamical systems and
  stochastic programming - from Ordinary Differential Equations and back}}.
\newblock {\sl \bibinfo{journal}{Transactions of Computational
  Systems Biology}, in print}.
\bibitemend

\bibitemstart{SB:Bortolussi:2007:QAPL:SPAandODE}
\bibinfo{author}{L.~Bortolussi} \& \bibinfo{author}{A.~Policriti}
  (\bibinfo{year}{2007}): \emph{\bibinfo{title}{Stochastic Concurrent
  Constraint Programming and Differential Equations}}.
\newblock In: {\sl \bibinfo{booktitle}{Proceedings of Fifth Workshop on
  Quantitative Aspects of Programming Languages, QAPL 2007}}, {\sl
  \bibinfo{series}{ENTCS}} \bibinfo{volume}{167}.
\bibitemend

\bibitemstart{SB:Bortolussi:2008:BiomodelingSCCP:Journal}
\bibinfo{author}{L.~Bortolussi} \& \bibinfo{author}{A.~Policriti}
  (\bibinfo{year}{2008}): \emph{\bibinfo{title}{Modeling Biological Systems in
  Concurrent Constraint Programming}}.
\newblock {\sl \bibinfo{journal}{Constraints}}
  \bibinfo{volume}{13}(\bibinfo{number}{1}).
\bibitemend

\bibitemstart{PA:Bortolussi:2009:HybridPIJournal}
\bibinfo{author}{L.~Bortolussi} \& \bibinfo{author}{A.~Policriti}
  (\bibinfo{year}{2009}): \emph{\bibinfo{title}{Hybrid Dynamics of Stochastic
  $\pi$-calculus}}.
\newblock {\sl \bibinfo{journal}{Mathematics in Computer Science}}
  \bibinfo{volume}{2}(\bibinfo{number}{3}), pp. \bibinfo{pages}{465--491}.
\bibitemend

\bibitemstart{SB:Bortolussi:2009:HybridSCCPstaticJournal}
\bibinfo{author}{L.~Bortolussi} \& \bibinfo{author}{A.~Policriti}
  (\bibinfo{year}{2009}): \emph{\bibinfo{title}{Hybrid Dynamics of Stochastic
  Programs}}.
\newblock {\sl \bibinfo{journal}{Submitted to Theor. Comp. Sc.}} .
\bibitemend

\bibitemstart{SB:Bortolussi:2009:HybridsCCPLattice}
\bibinfo{author}{L.~Bortolussi} \& \bibinfo{author}{A.~Policriti}
  (\bibinfo{year}{2009}): \emph{\bibinfo{title}{Stochastic Programs and Hybrid
  Automata for (Biological) Modeling}}.
\newblock In: {\sl \bibinfo{booktitle}{Proceedings of CiE 2009}}.
\bibitemend

\bibitemstart{HA:BujorianuLygeros:2004:GSHA}
\bibinfo{author}{M.L. Bujorianu} \& \bibinfo{author}{J.~Lygeros}
  (\bibinfo{year}{2004}): \emph{\bibinfo{title}{General Stochastic Hybrid
  Systems: Modeling and Optimal Control}}.
\newblock In: {\sl \bibinfo{booktitle}{Proceedings of 43rd IEEE Conference on
  Decision and Control (CDC 2004)}}. pp. \bibinfo{pages}{182--187}.
\bibitemend

\bibitemstart{SB:HillstonGilmoreCalder:2006:ERKPathwayPEPA}
\bibinfo{author}{M.~Calder}, \bibinfo{author}{S.~Gilmore} \&
  \bibinfo{author}{J.~Hillston} (\bibinfo{year}{2006}):
  \emph{\bibinfo{title}{Modelling the influence of RKIP on the ERK signalling
  pathway using the stochastic process algebra {PEPA}}}.
\newblock {\sl \bibinfo{journal}{Transactions on Computational Systems
  Biology}} \bibinfo{volume}{4230}, pp. \bibinfo{pages}{1--23}.
\bibitemend

\bibitemstart{STOC:Davis:1993:PDMP}
\bibinfo{author}{M.H.A. Davis} (\bibinfo{year}{1993}):
  \emph{\bibinfo{title}{Markov Models and Optimization}}.
\newblock \bibinfo{publisher}{Chapman \& Hall}.
\bibitemend

\bibitemstart{SB:Gillespie:2000:ChemicalLangevinEquation}
\bibinfo{author}{D.~Gillespie} (\bibinfo{year}{2000}):
  \emph{\bibinfo{title}{The chemical Langevin equation}}.
\newblock {\sl \bibinfo{journal}{Journal of Chemical Physics}}
  \bibinfo{volume}{113}(\bibinfo{number}{1}), pp. \bibinfo{pages}{297--306}.
\bibitemend

\bibitemstart{SB:Gillespie:1976:gillespieAlgorithm}
\bibinfo{author}{D.T. Gillespie} (\bibinfo{year}{1976}):
  \emph{\bibinfo{title}{A General Method for Numerically Simulating the
  Stochastic Time Evolution of Coupled Chemical Reactions}}.
\newblock {\sl \bibinfo{journal}{J. of Computational Physics}}
  \bibinfo{volume}{22}.
\bibitemend

\bibitemstart{SB:Gillespie:1977:gillespieAlgorithm}
\bibinfo{author}{D.T. Gillespie} (\bibinfo{year}{1977}):
  \emph{\bibinfo{title}{Exact Stochastic Simulation of Coupled Chemical
  Reactions}}.
\newblock {\sl \bibinfo{journal}{J. of Physical Chemistry}}
  \bibinfo{volume}{81}(\bibinfo{number}{25}).
\bibitemend

\bibitemstart{SB:Griffith:2006:HybridSimulation}
\bibinfo{author}{M.~Griffith}, \bibinfo{author}{T.~Courtney},
  \bibinfo{author}{J.~Peccoud} \& \bibinfo{author}{W.H. Sanders}
  (\bibinfo{year}{2006}): \emph{\bibinfo{title}{Dynamic partitioning for hybrid
  simulation of the bistable HIV-1 transactivation network}}.
\newblock {\sl \bibinfo{journal}{Bioinformatics}}
  \bibinfo{volume}{22}(\bibinfo{number}{22}), pp. \bibinfo{pages}{2782--2789}.
\bibitemend

\bibitemstart{PA:Gupta:1999:PCCP}
\bibinfo{author}{V.~Gupta}, \bibinfo{author}{R.~Jagadeesan} \&
  \bibinfo{author}{P.~Panangaden} (\bibinfo{year}{1999}):
  \emph{\bibinfo{title}{Stochastic processes as concurrent constraint
  programs}}.
\newblock In: {\sl \bibinfo{booktitle}{Proceedings of POPL'99}}.
\bibitemend

\bibitemstart{PA:Gupta:1997:PCCP}
\bibinfo{author}{V.~Gupta}, \bibinfo{author}{R.~Jagadeesan} \&
  \bibinfo{author}{V.A. Saraswat} (\bibinfo{year}{1997}):
  \emph{\bibinfo{title}{Probabilistic Concurrent Constraint Programming}}.
\newblock In: {\sl \bibinfo{booktitle}{Proceedings of CONCUR'97}}.
\bibitemend

\bibitemstart{SB:Rawlings:2002:HybridSimulation}
\bibinfo{author}{E.L. Haseltine} \& \bibinfo{author}{J.B. Rawlings}
  (\bibinfo{year}{2002}): \emph{\bibinfo{title}{Approximate simulation of
  coupled fast and slow reactions for stochastic chemical kinetics}}.
\newblock {\sl \bibinfo{journal}{Journal of Chemical Physics}}
  \bibinfo{volume}{117}(\bibinfo{number}{15}).
\bibitemend

\bibitemstart{SB:HaseltineRawlings:2005:OriginsApproximationChemKin}
\bibinfo{author}{E.L. Haseltine} \& \bibinfo{author}{J.B. Rawlings}
  (\bibinfo{year}{2005}): \emph{\bibinfo{title}{On the origins of
  approximations for stochastic chemical kinetics}}.
\newblock {\sl \bibinfo{journal}{J. Chem. Phys.}} \bibinfo{volume}{123}.
\bibitemend

\bibitemstart{HA:Henziger:1996:SurveyHybridAutomata}
\bibinfo{author}{T.~A. Henzinger} (\bibinfo{year}{1996}):
  \emph{\bibinfo{title}{The theory of hybrid automata}}.
\newblock In: {\sl \bibinfo{booktitle}{LICS '96: Proceedings of the 11th Annual
  IEEE Symposium on Logic in Computer Science}}.
\bibitemend

\bibitemstart{PA:Hillston:2005:ODEandPEPA}
\bibinfo{author}{J.~Hillston} (\bibinfo{year}{2005}):
  \emph{\bibinfo{title}{Fluid Flow Approximation of {PEPA} models}}.
\newblock In: {\sl \bibinfo{booktitle}{Proceedings of the Second International
  Conference on the Quantitative Evaluation of Systems (QEST�05)}}.
\bibitemend

\bibitemstart{SB:Kiehl:2004:hybridSimulation}
\bibinfo{author}{T.R. Kiehl}, \bibinfo{author}{R.M. Mattheyses} \&
  \bibinfo{author}{M.K. Simmons} (\bibinfo{year}{2004}):
  \emph{\bibinfo{title}{Hybrid Simulation of Cellular Behavior}}.
\newblock {\sl \bibinfo{journal}{Bioinformatics}}
  \bibinfo{volume}{20}(\bibinfo{number}{3}), pp. \bibinfo{pages}{316--322}.
\bibitemend

\bibitemstart{SB:Neogi:2004:HybridSimulationDynamicPartitioning}
\bibinfo{author}{N.~A. Neogi} (\bibinfo{year}{2004}):
  \emph{\bibinfo{title}{Dynamic Partitioning of Large Discrete Event Biological
  Systems for Hybrid Simulation and Analysis}}.
\newblock In: {\sl \bibinfo{booktitle}{Proceedings of 7th International
  Workshop on Hybrid Systems: Computation and Control, HSCC 2004}}, {\sl
  \bibinfo{series}{LNCS}} \bibinfo{volume}{2993}. pp.
  \bibinfo{pages}{463--476}.
\bibitemend

\bibitemstart{SB:Pahle:2009:HybridSimSurvey}
\bibinfo{author}{J.~Pahle} (\bibinfo{year}{2009}):
  \emph{\bibinfo{title}{Biochemical simulations: stochastic, approximate
  stochastic and hybrid approaches}}.
\newblock {\sl \bibinfo{journal}{Brief Bioinform.}}
  \bibinfo{volume}{10}(\bibinfo{number}{1}), pp. \bibinfo{pages}{53--64}.
\bibitemend

\bibitemstart{PA:DiPierro:1998:pCCP}
\bibinfo{author}{A.~Di Pierro} \& \bibinfo{author}{H.~Wiklicky}
  (\bibinfo{year}{1998}): \emph{\bibinfo{title}{An operational semantics for
  probabilistic concurrent constraint programming}}.
\newblock In: {\sl \bibinfo{booktitle}{Proceedings of IEEE Computer Society
  International Conference on Computer Languages}}.
\bibitemend

\bibitemstart{SB:ShapiroRegev:2002:CellsAsComputation}
\bibinfo{author}{A.~Regev} \& \bibinfo{author}{E.~Shapiro}
  (\bibinfo{year}{2002}): \emph{\bibinfo{title}{Cellular Abstractions: Cells as
  Computation}}.
\newblock {\sl \bibinfo{journal}{Nature}} \bibinfo{volume}{419}.
\bibitemend

\bibitemstart{SB:Kaznessis:2005:HybridSimulation}
\bibinfo{author}{H.~Salis} \& \bibinfo{author}{Y.~Kaznessis}
  (\bibinfo{year}{2005}): \emph{\bibinfo{title}{Accurate hybrid stochastic
  simulation of a system of coupled chemical or biochemical reactions}}.
\newblock {\sl \bibinfo{journal}{Journal of Chemical Physics}}
  \bibinfo{volume}{122}.
\bibitemend

\bibitemstart{PA:Saraswat:1990:CCP}
\bibinfo{author}{V.~Saraswat} \& \bibinfo{author}{M.~Rinard}
  (\bibinfo{year}{1990}): \emph{\bibinfo{title}{Concurrent Constraint
  Programming}}.
\newblock In: {\sl \bibinfo{booktitle}{Proceedings of 18th Symposium on
  Principles Of Programming Languages (POPL)}}.
\bibitemend

\bibitemstart{PA:Saraswat:1993:CCP}
\bibinfo{author}{V.~A. Saraswat} (\bibinfo{year}{1993}):
  \emph{\bibinfo{title}{Concurrent Constraint Programming}}.
\newblock \bibinfo{publisher}{MIT press}.
\bibitemend

\bibitemstart{HA:Strubbe:2007:CommunicatingPDP}
\bibinfo{author}{S.~Strubbe} \& \bibinfo{author}{A.~van~der Schaft}
  (\bibinfo{year}{2007}): \emph{\bibinfo{title}{Stochastic Hybrid Systems}},
  chapter \bibinfo{chapter}{Compositional Modeling of Stochastic Hybrid
  Systems}, pp. \bibinfo{pages}{47--78}.
\newblock \bibinfo{publisher}{CRC Press}.
\bibitemend

\bibitemstart{SB:Wilkinson:2006:StochasticModellingSB}
\bibinfo{author}{D.~J. Wilkinson} (\bibinfo{year}{2006}):
  \emph{\bibinfo{title}{Stochastic Modelling for Systems Biology}}.
\newblock \bibinfo{publisher}{Chapman \& Hall}.
\bibitemend

\bibliographyend
\end{thebibliography}


\end{document}